\documentclass[aps,pra,preprint,showpacs]{revtex4}
\usepackage{bm}
\newcommand{\phiel}{\phi_\mathrm{el}}
\begin{document}
\preprint{Version 1.0}

\title{Finite nuclear mass corrections to electric and 
       magnetic interactions in diatomic molecules}

\author{Krzysztof Pachucki}
\email[]{krp@fuw.edu.pl} \homepage[]{www.fuw.edu.pl/~krp}
\affiliation{Institute of Theoretical Physics, 
             University of Warsaw,
             Ho\.{z}a 69, 00-681 Warsaw, Poland}

\date{\today}

\begin{abstract}
In order to interpret precise measurements of molecular properties
the finite nuclear mass corrections to the Born-Oppenheimer approximation
have to be accounted for. It is shown that they can be obtained
systematically in the perturbative approach. The formulae for
the leading corrections to the relativistic contribution to energy, 
the transition electric dipole moment, the electric polarizability, 
and the magnetic shielding constant are presented.
\end{abstract}

 \maketitle

\section{Introduction}
In the calculation of some molecular property, the magnetic shielding
for example, one usually assumes fixed position of nuclei,
the so called Born-Oppenheimer (BO) approximation
and at the final stage, averages over the appropriate vibration-rotational
function. The principal question we address in this work
is, what are the finite nuclear mass corrections to various physical properties 
such as relativistic energies, polarizabilities or the mentioned magnetic
shielding evaluated in the BO approximation.
The direct nonadiabatic calculations are possible only for small
molecules and only for simple properties such as the electric dipole 
polarizability \cite{adam1, adam2}. 
This approach however is not universal and has not yet been applied
to more complex molecules such as H$_2$O, or to
the evaluation of the nuclear spin-rotation and shielding constants.
In this work we demonstrate the applicability of the nonadiabatic 
perturbation theory \cite{nonad} to obtain in a systematic way 
the formulae for various physical properties of a diatomic molecule, 
with possible extensions to larger molecules. 
We rederive the known result for the electric static polarizability, the
rotational magnetic moment, the spin-rotation constant, and obtain 
the leading finite nuclear mass corrections
(which we will call in this work nonadiabatic corrections)
to relativistic rovibrational energies, the transition electric dipole moment,
the electric static polarizability, and the magnetic shielding constant.
These finite nuclear mass corrections are important 
for comparison between accurate measurement and precise calculations,
for example in the dissociation energy of H$_2 $ \cite{merkt,jeziorsk}, 
the transition electric-dipole moment of LiH \cite{adam3}, or
in the shielding constant of H$_2$ and isotopomers \cite{gauss}.

We demonstrate in this work that the leading finite nuclear mass
corrections can be  conveniently calculated for a fixed position of nuclei, 
as in the BO approximation, and averaged out over
the rovibrational function. We do not consider here the 
second order corrections in the nuclear Hamiltonian $H_{\rm n}$,
but for consistency regard them also as the nonadiabatic corrections,
although of higher order.

In Sec. II we define the reference frame
and split the nonrelativistic Hamiltonian into the electronic and nuclear parts.
In Sec. III we briefly present nonadiabatic perturbation theory 
on the basis of Ref. \cite{nonad}, include nonadiabatic corrections to 
the BO wave function and derive general formulae for the first and the second order
matrix elements. In Sec. IV, as a first example, we derive formulae 
for relativistic recoil corrections to rovibrational energies 
in diatomic molecules. Other examples, the finite nuclear mass corrections
to electric properties of molecules are derived in Sec. V,
and to magnetic properties in Sec.~VI.
Finally, the obtained results are briefly summarized in Sec. VII.

\section{Nonrelativistic Hamiltonian}
We consider a neutral diatomic molecule with the Hamiltonian 
\begin{equation}
H = \sum_a\frac{\vec p_a^{\;2}}{2\,m} +\frac{\vec p_A^{\;2}}{2\,m_A}
+\frac{\vec p_B^{\;2}}{2\,m_B} + V, 
\label{01}
\end{equation}
where the summation index $a$ goes over all 
electrons and $A$ and $B$ refers to nuclei.
In order to derive formulae for nonadiabatic effects,
one must fix the reference frame. We start with the
laboratory frame $\{\vec R_A,\,\vec R_B,\,\vec r_a\}$, and 
change  variables to $\{\vec R,\,\vec R_G,\,\vec x_a\}$
according to
\begin{eqnarray}
\vec R_A &=& \vec R_G+\epsilon_B\,\vec R,\label{02}\\
\vec R_B &=& \vec R_G-\epsilon_A\,\vec R,\label{03}\\ 
\vec r_a &=& \vec R_G+\vec x_a, \label{04}
\end{eqnarray}
with  the relative position of nuclei
$\vec R = \vec R_A-\vec R_B$, and the arbitrarily chosen  
on the molecular symmetry axis new frame origin
$\vec R_G = \epsilon_A\,\vec R_A + \epsilon_B\,\vec R_B$, where
$\epsilon_A+\epsilon_B=1$.
The conjugate momenta are related by
\begin{eqnarray}
{\vec p}_A &=& \epsilon_A\,\vec P_G + {\vec P} -\epsilon_A\,\sum_a\vec q_a,\\
{\vec p}_B &=& \epsilon_B\,\vec P_G - {\vec P} -\epsilon_B\,\sum_a\vec q_a,\\
{\vec p}_a &=& \vec q_a,
\end{eqnarray}
where $\vec P = -i\,\vec\nabla_R$ and $\vec q_a =
-i\,\vec\nabla_{x_a}$. 
The nonrelativistic wave function 
with the vanishing total momentum does not depend on $\vec R_G$, 
thus $\phi = \phi(\vec x_a,\,\vec R)$ and since $\vec P_G$ commutes with $H$
when expressed in new variables, it can be set to $0$. 
Hamiltonian $H$ in new variables becomes
\begin{eqnarray}
H &=& \sum_a \frac{\vec q_a^{\;2}}{2\,m} + V 
+\biggl(\frac{1}{2\,m_A}+\frac{1}{2\,m_B}\biggr)\,\vec P^{\;2}
+\biggl(\frac{\epsilon_A^2}{2\,m_A}+\frac{\epsilon_B^2}{2\,m_B}\biggr)\,
\Bigl(\sum_a\vec q_a\Bigr)^2\nonumber \\ &&
-\biggl(\frac{\epsilon_A}{m_A}-\frac{\epsilon_B}{m_B}\biggr)\,
\vec P\cdot\sum_a\vec q_a.
\label{08}
\end{eqnarray}
The last term in the above is transformed by the unitary transformation 
\begin{equation}
\tilde H = e^{-i\,\varphi}\,H\,e^{i\,\varphi} = H -i [\varphi\;,\;H]+\ldots
\end{equation}
where
\begin{equation}
\varphi = m\,\biggl(\frac{\epsilon_A}{m_A}-\frac{\epsilon_B}{m_B}\biggr)\,
\sum_a\vec x_a\cdot\vec P, \label{09}
\end{equation}
and
\begin{equation}
e^{-i\,\varphi}\,q_a^i\,e^{i\,\varphi} = q_a^i 
+m\,\biggl(\frac{\epsilon_A}{m_A}-\frac{\epsilon_B}{m_B}\biggr)\,P^i,
\end{equation}
with higher order $O(m/M)^2$ terms in the electron nuclear mass ratio
being neglected and $M$ is equal to $m_A$ or $m_B$.
As a result of this transformation Hamiltonian takes the form
\begin{eqnarray}
\tilde H &=& H_{\rm el} + H_{\rm n},\label{12}\\
H_{\rm el} &=& \sum_a \frac{\vec q_a^{\;2}}{2\,m} + V, \\
H_{\rm n} &=& \biggl(\frac{1}{2\,m_A}+\frac{1}{2\,m_B}\biggr)\,\vec P^{\;2}
+\biggl(\frac{\epsilon_A^2}{2\,m_A}+\frac{\epsilon_B^2}{2\,m_B}\biggr)\,
\Bigl(\sum_a\vec q_a\Bigr)^2 \nonumber\\&&
-m\biggl(\frac{\epsilon_A}{m_A}-\frac{\epsilon_B}{m_B}\biggr)\,
\sum_a\vec x_a\cdot\vec\nabla_R(V)\label{14}\\ &=&
H_{\rm n}' + H_{\rm n}'' ,\label{15}
\end{eqnarray}
where $H_{\rm n}'$ includes the first term and $H_{\rm n}''$
the two remaining terms.
This form of the nuclear Hamiltonian is convenient for the
calculation of nonadiabatic effects. Moreover the freedom in
choosing $\epsilon_{A,B}$ will be used in order to simplify 
formulae for nonadiabatic corrections to electric and magnetic properties. 
Later, we will need the angular momentum operator $\vec J$,
which for states with the vanishing total momentum is defined by
\begin{equation}
\vec J = \sum_a (\vec r_a-\vec R_{CM})\times\vec p_a + 
         (\vec R_A -\vec R_{CM})\times\vec p_A +
         (\vec R_B -\vec R_{CM})\times\vec p_B,
\label{16}
\end{equation}
where $\vec R_{CM}$ is the molecular mass center. In new variables
the operator $\vec J$ 
\begin{equation}
\vec J = \sum_a \vec x_a\times\vec q_a +\vec R\times\vec P 
\equiv \vec J_{\rm el}+\vec J_{\rm n}
\label{17}
\end{equation}
is split into electronic $\vec J_{\rm el}$ and nuclear $\vec J_{\rm n}$ 
parts, and this $\vec J$
is not being modified by the unitary transformation of Eq. (\ref{09}).

\section{Nonadiabatic perturbation theory}
The total nonrelativistic wave function $\phi$ of an arbitrary molecule
is the solution of the stationary Schr\"odinger equation 
\begin{equation}
[H-E]\,|\phi\rangle = 0\,, \label{18}
\end{equation}
with the Hamiltonian $H$ being a sum of the
electronic $H_{\rm el}$ and nuclear $H_{\rm n}$ parts, Eq. (\ref{12}).
In the adiabatic approximation $\phi=\phi_{\rm a}$, where
\begin{equation}
\phi_{\rm a}(\vec x,\vec R) = \phi_{\rm el}(\vec x) \; \chi(\vec R) \label{19}
\end{equation}
is represented as a product of the electronic wave function $\phi_{\rm el}$
and the nuclear wave function $\chi$. We note, that $\phi_{\rm el}$
depends implicitly on the nuclear relative coordinate $\vec R$.
The electronic wave function obeys the clamped nuclei electronic 
Schr\"odinger equation 
\begin{equation}
\bigl[H_{\rm el}-{\cal E}_{\rm el}(R)\bigr]\,|\phi_{\rm el}\rangle = 0, \label{20}
\end{equation} 
while the nuclear wave function is a solution to the Schr\"odinger equation 
in the effective potential generated by electrons
\begin{equation}
\bigl[ H_{\rm n} +{\cal E}_{\rm a}(R)+{\cal E}_{\rm el}(R)-E_{\rm a}\bigr]\,
|\chi\rangle = 0\,, \label{21}
\end{equation} 
where 
\begin{equation}
{\cal E}_{\rm a}(R) = \bigl\langle\phiel|H_{\rm n}|\phiel\bigr\rangle_{\rm el}\,.
\label{22}
\end{equation}

In the nonadiabatic perturbation theory, 
the total wave function
\begin{equation}
\phi = \phi_{\rm a} + \delta\phi_{\rm na} = \phi_{\rm el}\,\chi + \delta\phi_{\rm na}
\label{23}
\end{equation}
is the sum of the adiabatic solution and a nonadiabatic correction. 
The nonadiabatic correction $\delta\phi_{\rm na}$ is  decomposed into two parts
\begin{equation}
\delta\phi_{\rm na} = \phi_{\rm el}\,\delta\chi + \delta'\phi_{\rm na}\,,
\label{24}
\end{equation}
obeying the following orthogonality conditions
\begin{eqnarray}
\langle\delta'\phi_{\rm na}|\phi_{\rm el}\rangle_{\rm el} &=& 0\,,\label{25}\\
\langle\delta\chi|\chi\rangle &=& 0\,,\label{26} 
\end{eqnarray}
which imply the normalization condition $\langle\phi_a|\phi\rangle=1$.

In the first order in $H_{\rm n}$ of the nonadiabatic perturbation theory
one has
\begin{eqnarray}
|\delta'\phi_{\rm na}\rangle^{(1)} &=& 
\frac{1}{({\cal E}_{\rm el}-H_{\rm el})'}\,
H_{\rm n}\,|\phi_{\rm el}\,\chi\rangle,
\label{27}
\end{eqnarray}
and in the second order
\begin{eqnarray}
|\delta'\phi_{\rm na}\rangle^{(2)} &=& 
\frac{1}{({\cal E}_{\rm el}-H_{\rm el})'}\,
(H_{\rm n}+{\cal E}_{\rm el}-E_{\rm a})
\,\frac{1}{({\cal E}_{\rm el}-H_{\rm el})'}\,
H_{\rm n}\,|\phi_{\rm el}\,\chi\rangle,
\label{28}
\end{eqnarray}
where $1/({\cal E}_{\rm el}-H_{\rm el})'$ denotes resolvent with
the reference state $\phi_{\rm el}$ subtracted out.
The total nuclear function $\chi+\delta\chi$ satisfies 
the effective Schr\"odinger equation which includes adiabatic and 
nonadiabatic corrections \cite{nonad}. Thus the
nonadiabatic wave function can be recovered order by order in the perturbative approach.
Eqs. (\ref{27},\ref{28}) involve $H_{\rm n}$, and thus derivatives 
with respect to $\vec R$. These derivatives can be calculated with the help
of the following formulae
\begin{eqnarray}
\vec\nabla_R |\phi_{\rm el}\rangle &=&
\frac{1}{({\cal E}_{\rm el}-H_{\rm el})'}
\,\vec\nabla_R(V)|\phi_{\rm el}\rangle,\label{29}\\
\vec\nabla_R\biggl[\frac{1}{({\cal E}_{\rm el}-H_{\rm el})'}\biggr] &=& 
\frac{1}{({\cal E}_{\rm el}-H_{\rm el})'}
\,\vec\nabla_R(V-{\cal E}_{\rm el})
\,\frac{1}{({\cal E}_{\rm el}-H_{\rm el})'}\nonumber \\ &&
-\frac{1}{({\cal E}_{\rm el}-H_{\rm el})'^2}
\,\vec\nabla_R(V)
|\phi_{\rm el}\rangle\langle\phi_{\rm el}|\nonumber \\ &&
-|\phi_{\rm el}\rangle\langle\phi_{\rm el}|
\,\vec\nabla_R(V)
\,\frac{1}{({\cal E}_{\rm el}-H_{\rm el})'^2}.\label{30}
\end{eqnarray}
It has been shown recently in \cite{nonad}, that the application
of these formulae allows for a significant improvement
in the numerical accuracy of adiabatic
and nonadiabatic corrections in H$_2$ molecule, and
this probably will hold for any diatomic molecule.
Alternatively, one may use the formula
\begin{equation}
\vec\nabla_R = \vec n\,(\vec n\cdot\vec\nabla_R) 
-\vec n\times(\vec n\times\vec\nabla_R),\label{31}
\end{equation}
where $\vec n=\vec R/R$, with only the first radial part in above equation 
replaced in terms of the derivative 
$\partial(V-{\cal E}_{\rm el})/\partial R$, namely
\begin{equation}
\vec\nabla_R |\phi_{\rm el}\rangle =\vec n\,
\frac{1}{({\cal E}_{\rm el}-H_{\rm el})'}
\,\frac{\partial V}{\partial R}|\phi_{\rm el}\rangle
-\frac{i}{R}\vec n\times\vec J_{\rm n}|\phi_{\rm el}\rangle.
\label{32}
\end{equation}
For example the adiabatic correction to energy becomes
\begin{eqnarray}
{\cal E}_{\rm a}(R) &=& \langle\phi_{\rm el}|H_{\rm n}|\phi_{\rm el}\rangle
\nonumber\\
&=& \langle\phi_{\rm el}|H''_{\rm n}+\frac{\vec J_{\rm n}^2}{2\,m_{\rm n}\,R^2}|\phi_{\rm el}\rangle
+\langle\phi_{\rm el}|\frac{\partial V}{\partial R}\,
\frac{1}{({\cal E}_{\rm el}-H_{\rm el})'^2}\frac{\partial V}{\partial R}
|\phi_{\rm el}\rangle,
\end{eqnarray}
where $m_{\rm n}$ is the nuclear reduced mass
and for $\Sigma$ electronic state $\vec J_{\rm n}|\phi_{\rm el}\rangle$ 
can be replaced by $-\vec J_{\rm el}|\phi_{\rm el}\rangle$.
In this way one avoids summation over intermediate states with the $\Pi$ symmetry.

\subsection{First order matrix elements}
We will use here this nonadiabatic perturbation theory 
to derive the finite nuclear mass corrections to various matrix elements
in the general form. Later we will analyze specific examples. 
Consider the Hermitian electronic operator $Q$ (no derivatives with respect
to nuclear variables, for example the relativistic correction to kinetic energy of
electrons) and its matrix element between 
(different) rovibrational states.
In the BO approximation this matrix element can be represented 
in terms of the electronic matrix element nested in the nuclear matrix element, namely
\begin{eqnarray}
\langle Q\rangle^{(0)} &\equiv& 
\langle\phi_{\rm el}\,\chi_{\rm f}|Q|\phi_{\rm el}\,\chi_{\rm i}\rangle 
= \langle\chi_{\rm f}|\langle Q\rangle_{\rm el}^{(0)}|\chi_{\rm i}\rangle, \\
\langle Q\rangle_{\rm el}^{(0)} &\equiv& \langle Q\rangle_{\rm el}  
= \langle\phi_{\rm el}|Q|\phi_{\rm el}\rangle.
\label{35}
\end{eqnarray}
We will show  that the same holds for 
nonadiabatic corrections to this matrix element, which are
\begin{equation}
\langle Q\rangle^{(1)} = 
\langle\phi_{\rm el}\,\chi_{\rm f}|H_{\rm n}\,
\frac{1}{({\cal E}_{\rm el}-H_{\rm el})'}\,
Q|\phi_{\rm el}\,\chi_{\rm i}\rangle +
\langle\phi_{\rm el}\,\chi_{\rm f}|Q\,
\frac{1}{({\cal E}_{\rm el}-H_{\rm el})'}\,
H_{\rm n}|\phi_{\rm el}\,\chi_{\rm i}\rangle.
\label{36}
\end{equation}
Although in this work we consider only the first order $O(m/M)$
corrections, let us present here the second order corrections to 
the diagonal matrix element to demonstrate the application of 
the nonadiabatic perturbation theory
\begin{eqnarray}
\langle Q\rangle^{(2)} &=& 
\langle\phi_{\rm el}\,\chi|H_{\rm n}\,
\frac{1}{({\cal E}_{\rm el}-H_{\rm el})'}\,
(H_{\rm n}+{\cal E}_{\rm el}-E_{\rm a})
\,\frac{1}{({\cal E}_{\rm el}-H_{\rm el})'}\,
Q|\phi_{\rm el}\,\chi\rangle 
\nonumber \\ &&
+\langle\phi_{\rm el}\,\chi|Q\,
\frac{1}{({\cal E}_{\rm el}-H_{\rm el})'}\,
(H_{\rm n}+{\cal E}_{\rm el}-E_{\rm a})
\,\frac{1}{({\cal E}_{\rm el}-H_{\rm el})'}\,
H_{\rm n}|\phi_{\rm el}\,\chi\rangle
\nonumber \\ &&
+\langle\phi_{\rm el}\,\chi|H_{\rm n}\,
\frac{1}{({\cal E}_{\rm el}-H_{\rm el})'}\,
(Q-\langle Q \rangle^{(0)})\,\frac{1}{({\cal E}_{\rm el}-H_{\rm el})'}\,
H_{\rm n}|\phi_{\rm el}\,\chi\rangle. \label{37}
\end{eqnarray}
Additional corrections due to $\delta\chi$ in Eq.(\ref{24}) can easily be included
in the $\langle Q \rangle^{(0)}$ and $\langle Q \rangle^{(1)}$
by replacing $\chi$ by $\chi+\delta\chi$
and will not be considered any further.  
Let us return now to the leading order correction to the matrix element in
Eq. (\ref{36}) 
\begin{eqnarray}
\langle Q\rangle^{(1)} &=& 
\int d^3R\,\biggl\{(\chi^*_{\rm f}\,\chi_{\rm i})\,\biggl[
\langle H_{\rm n}\,\phi_{\rm el}|
\frac{1}{({\cal E}_{\rm el}-H_{\rm el})'}\,Q\,|\phi_{\rm el}\rangle
+\langle\phi_{\rm el}|Q\,\frac{1}{({\cal E}_{\rm el}-H_{\rm el})'}\,
|H_{\rm n}\,\phi_{\rm el}\rangle\biggr]
\nonumber \\ &-&
\frac{
\vec\nabla\bigl(\chi^*_{\rm f}\,\chi_{\rm i}\bigr)}{2\,m_{\rm n}}\,
\biggl[
\langle\vec\nabla_R\phi_{\rm el}|
\frac{1}{({\cal E}_{\rm el}-H_{\rm el})'}\,Q\,|\phi_{\rm el}\rangle
+\langle\phi_{\rm el}|Q\,\frac{1}{({\cal E}_{\rm el}-H_{\rm el})'}\,
|\vec\nabla_R\phi_{\rm el}\rangle\biggr]
\label{38}\\ &-&
\frac{
\Bigl(\chi_{\rm i}\,\vec\nabla\chi^*_{\rm f}-
\chi_{\rm f}^*\,\vec\nabla\chi_{\rm i}\Bigr)}{2\,m_{\rm n}}\,
\biggl[
\langle\vec\nabla_R\phi_{\rm el}|
\frac{1}{({\cal E}_{\rm el}-H_{\rm el})'}\,Q\,|\phi_{\rm el}\rangle
-\langle\phi_{\rm el}|Q\,\frac{1}{({\cal E}_{\rm el}-H_{\rm el})'}\,
|\vec\nabla_R\phi_{\rm el}\rangle\biggr]\biggr\}\nonumber
\end{eqnarray}
and consider two special cases. If $Q$ is a real operator, 
then the third term vanishes and with the help of integration by parts
we obtain
\begin{eqnarray}
\langle Q \rangle^{(1)} &=&\langle\chi_{\rm f}|
           \langle Q \rangle^{(1)}_{\rm el} |\chi_{\rm i}\rangle, \label{39}\\
\langle Q \rangle^{(1)}_{\rm el} &=& \langle\phi_{\rm el}|
\stackrel{\leftrightarrow}{H_{\rm n}}
\frac{1}{({\cal E}_{\rm el}-H_{\rm el})'} \,Q\, |\phi_{\rm el}\rangle + 
\langle\phi_{\rm el}| \,Q\, \frac{1}{({\cal E}_{\rm el}-H_{\rm el})'}
\stackrel{\leftrightarrow}{H_{\rm n}}|\phi_{\rm el}\rangle,
\label{40}
\end{eqnarray}
where for arbitrary $\psi_{\rm el}$ and $\psi'_{\rm el}$:
\begin{equation}
\langle\psi'_{\rm el}|\!\!\stackrel{\leftrightarrow}{H_{\rm n}}\!\!|
\psi_{\rm el}\rangle
 = \langle\vec\nabla_R\,\psi'_{\rm el}|\vec\nabla_R\,\psi_{\rm  el}\rangle/(2\,m_{\rm n})
+\langle\psi'_{\rm el}|H''_{\rm n}|\psi_{\rm el}\rangle.
\end{equation}
This case of the real Hermitian $Q$ finds applications in studying
relativistic corrections to rovibrational energies and to all
electric properties. 
If $Q = \vec Q$ is an imaginary vector operator 
composed of electronic operators and $\vec R$,
such that $\vec R\cdot\vec Q = 0$, then
the first two terms in Eq. (\ref{38}) vanish and 
\begin{eqnarray}
\langle Q^i \rangle^{(1)} &=& \frac{1}{2\,m_{\rm n}}\,
\int \frac{d^3 R}{R^2}\,[(J_{\rm n}^i\,\chi_{\rm f})^*\,\chi_{\rm i} 
+ \chi_{\rm f}^*\,(J_{\rm n}^i\,\chi_{\rm i})]\nonumber \\ &&
\times\biggl[\langle\phi_{\rm el}|Q^j\,
\frac{1}{({\cal E}_{\rm el}-H_{\rm el})'} 
|J_{\rm n}^j\phi_{\rm el}\rangle +
\langle J_{\rm n}^j\phi_{\rm el}|
\frac{1}{({\cal E}_{\rm el}-H_{\rm el})'}\,Q^j|\phi_{\rm el}\rangle\biggr],
\end{eqnarray}
where $\vec J_{\rm n}$ is defined in Eq. (\ref{17}). 
Let us assume that $\phi_{\rm el}$ is a $\Sigma$ state, 
then $\vec J_{\rm n}\,|\phi_{\rm el}\rangle = -\vec J_{\rm el}\,|\phi_{\rm el}\rangle$
and
\begin{equation}
\langle Q^i \rangle^{(1)} =
-\biggl\langle\chi_{\rm f}\biggl|\,\frac{J_{\rm n}^i}{m_{\rm n}\,R^2}\
\langle\phi_{\rm el}|J^j_{\rm el}\,
\frac{1}{({\cal E}_{\rm el}-H_{\rm el})'}\,
Q^j|\phi_{\rm el}\rangle\,\biggr|\chi_{\rm i}\biggr\rangle.
\label{43}
\end{equation}
This case of imaginary Hermitian $Q$ finds application in studying
the magnetic properties.

\subsection{Second order matrix elements}
Consider the second order matrix element with two arbitrary electronic
operators $Q_1$ and $Q_2$, Let us assume that 
$\langle\phi_{\rm el}|Q_i|\phi_{\rm el}\rangle=0$
and introduce the notation
\begin{equation}
\langle Q_1\;Q_2\rangle \equiv
\langle\phi|Q_1\,\frac{1}{(E-H)'}\,Q_2|\phi\rangle+{\rm c.c.}
\label{44}
\end{equation}
In the leading order of the nonadiabatic perturbation theory 
this matrix element is
\begin{eqnarray}
\langle Q_1\;Q_2\rangle^{(0)}
&=&
\langle\chi|\langle Q_1\;Q_2\rangle^{(0)}_{\rm el}|\chi\rangle, \\
\langle Q_1\;Q_2\rangle^{(0)}_{\rm el} &=& 
\;\langle\phi_{\rm el}|Q_1\,
\frac{1}{({\cal E}_{\rm el}-H_{\rm el})'}\,
Q_2|\phi_{\rm el}\rangle+ {\rm c.c.},
\label{46}
\end{eqnarray}
and the nonadiabatic correction is
\begin{eqnarray}
\langle Q_1\;Q_2\rangle^{(1)}
&=&
\langle\chi\,\phi_{\rm el}|Q_1\,
\frac{1}{({\cal E}_{\rm el}-H_{\rm el})'}\,
(H_{\rm n}+{\cal E}_{\rm el}-E_{\rm a})\,
\frac{1}{({\cal E}_{\rm el}-H_{\rm el})'}\,
Q_2|\phi_{\rm el}\,\chi\rangle\nonumber \\ &&
+\langle\chi\,\phi_{\rm el}|H_{\rm n}\,
\frac{1}{({\cal E}_{\rm el}-H_{\rm el})'}\,Q_1
\frac{1}{({\cal E}_{\rm el}-H_{\rm el})'}\,
Q_2|\phi_{\rm el}\,\chi\rangle\nonumber\\ &&
+\langle\chi\,\phi_{\rm el}|Q_1\,
\frac{1}{({\cal E}_{\rm el}-H_{\rm el})'}\,Q_2
\frac{1}{({\cal E}_{\rm el}-H_{\rm el})'}\,
H_{\rm n}|\phi_{\rm el}\,\chi\rangle + {\rm c.c.}\,.
\label{47}
\end{eqnarray}
This correction can also be rewritten in terms of 
the nested electronic matrix element, namely
\begin{eqnarray}
\langle Q_1\;Q_2\rangle^{(1)}
&=&\langle\chi|\langle Q_1\;Q_2\rangle^{(1)}_{\rm el}|\chi\rangle\\
\langle Q_1\;Q_2\rangle^{(1)}_{\rm el} &=&
\langle\phi_{\rm el}|Q_1\,
\frac{1}{({\cal E}_{\rm el}-H_{\rm el})'}\,
(\stackrel{\leftrightarrow}{H_{\rm n}}-{\cal E}_{\rm a})\,
\frac{1}{({\cal E}_{\rm el}-H_{\rm el})'}\,
Q_2|\phi_{\rm el}\rangle\nonumber \\ &&
+\langle\phi_{\rm el}|\stackrel{\leftrightarrow}{H_{\rm n}}\,
\frac{1}{({\cal E}_{\rm el}-H_{\rm el})'}\,Q_1
\frac{1}{({\cal E}_{\rm el}-H_{\rm el})'}\,
Q_2|\phi_{\rm el}\rangle\nonumber\\ &&
+\langle\phi_{\rm el}|Q_1\,
\frac{1}{({\cal E}_{\rm el}-H_{\rm el})'}\,Q_2
\frac{1}{({\cal E}_{\rm el}-H_{\rm el})'}\,
\stackrel{\leftrightarrow}{H_{\rm n}}|\phi_{\rm el}\rangle + {\rm c.c.}\,.
\label{49}
\end{eqnarray}
These formulae will be used in the calculations of the 
nonadiabatic corrections to the shielding constant.
The more general case with $\langle \phi_{\rm el}|Q_i|\phi_{\rm el}\rangle \neq 0$
for Hermitian real operators $Q_i$ is considered 
in the following section using a slightly different approach.

\subsection{Diagonal matrix elements with real Hermitian operators}
The finite nuclear mass corrections to the diagonal matrix element 
of a Hermitian and real operator $Q$ can obtained
by taking a derivative $\delta_Q$ with respect to $Q$ of 
the nuclear Schr\"odinger equation, 
which includes the diagonal adiabatic correction 
${\cal E}_{\rm a}$ in Eq. (\ref{21})
\begin{eqnarray}
\delta_Q\bigl[ H_{\rm n} +{\cal E}_{\rm a}(R)
+{\cal E}_{\rm el}(R)-E_{\rm a}\bigr]\,
|\chi\rangle = 0\,,
\label{50}
\end{eqnarray}
that is
\begin{eqnarray}
\bigl[ H_{\rm n} +{\cal E}_{\rm a}(R)+{\cal E}_{\rm el}(R)-E_{\rm a}\bigr]\,
|\delta_Q\chi\rangle +
\bigl[ \delta_Q{\cal E}_{\rm a}(R)+\delta_Q{\cal E}_{\rm el}(R)-\delta_QE_{\rm a}\bigr]\,
|\chi\rangle= 0\,.
\label{51}
\end{eqnarray}
Taking the product with $\langle\chi|$ on the left hand side, one obtains
the matrix element with the leading finite nuclear mass corrections
\begin{equation}
\langle\phi|Q|\phi\rangle 
=\delta_Q E \approx \delta_Q E_{\rm a} =
\langle\chi|\delta_Q{\cal E}_{\rm el}(R) + \delta_Q{\cal E}_{\rm a}(R)|\chi\rangle.
\end{equation}
The perturbation of electronic energies ${\cal E}_{\rm el}(R)$ 
and ${\cal E}_{\rm a}(R)$ due to some operator $Q$ can be obtained using
the standard Rayleigh-Schr\"odinger perturbation theory and the result
\begin{equation}
\langle\phi|Q|\phi\rangle =
\langle\chi|\langle Q\rangle_{\rm el}^{(0)}|\chi\rangle
+\langle\chi|\langle Q\rangle_{\rm el}^{(1)}|\chi\rangle + \ldots
\end{equation}
coincides with the former derivation. The fact that leading finite nuclear
mass corrections to the matrix elements can be obtained from the adiabatic
nuclear equation simplifies their derivation.

For the corrections to the second order matrix element we will need
$\delta_Q|\chi\rangle$ which is
\begin{eqnarray}
\delta_Q|\chi\rangle &=& \frac{1}
{\bigl[ E_{\rm a} - H_{\rm n} - {\cal E}_{\rm a}(R) - {\cal E}_{\rm el}(R)\bigr]'}\,
\Bigl[
\langle\phiel|Q|\phiel\rangle\nonumber \\ &&
+\langle\phiel|H_{\rm n}\,\frac{1}{({\cal E}_{\rm el}-H_{\rm el})'}\,Q|\phiel\rangle
+\langle\phiel|Q\,\frac{1}{({\cal E}_{\rm el}-H_{\rm el})'}\,H_{\rm n}|\phiel\rangle
\Bigr]\,|\chi\rangle. 
\label{54}
\end{eqnarray} 
Consider now the second order matrix element $\langle Q_1\;Q_2\rangle$ in
Eq. (\ref{44}) with two electronic operators $Q_1$ and $Q_2$.
In order to find the Born-Oppenheimer form and the finite nuclear mass corrections,
we take the second order derivative $\delta_{Q_1 Q_2}$ of Eq. (\ref{21}), 
and multiply from the left by $\langle\chi|$
\begin{eqnarray}
&&\langle\chi|\delta_{Q_1 Q_2}{\cal E}_{\rm a}(R)
+\delta_{Q_1 Q_2}{\cal E}_{\rm el}(R)-\delta_{Q_1 Q_2}E_{\rm a}\,
|\chi\rangle 
+\langle\chi|\delta_{Q_1}{\cal E}_{\rm a}(R)+\delta_{Q_1}{\cal E}_{\rm el}(R)\,
|\delta_{Q_2}\chi\rangle \nonumber \\&&
+\langle\chi|\delta_{Q_2}{\cal E}_{\rm a}(R)+\delta_{Q_2}{\cal E}_{\rm el}(R)\,
|\delta_{Q_1}\chi\rangle= 0\,.
\end{eqnarray}
This second order matrix element $\langle Q_1\;Q_2\rangle$
is identified with $\delta_{Q_1 Q_2}E_{\rm a}$, and thus
\begin{eqnarray}
\langle Q_1\;Q_2\rangle
&=&\langle\phi|Q_1\,\frac{1}{(E-H)'}\,Q_2|\phi\rangle +
\langle\phi|Q_2\,\frac{1}{(E-H)'}\,Q_1|\phi\rangle 
\nonumber\\ &=&
\langle\chi|\delta_{Q_1 Q_2}{\cal E}_{\rm el}(R)\,|\chi\rangle 
+\langle\chi|\delta_{Q_1}{\cal E}_{\rm el}(R)\,|\delta_{Q_2}\chi\rangle
+\langle\chi|\delta_{Q_2}{\cal E}_{\rm el}(R)\,|\delta_{Q_1}\chi\rangle\nonumber \\ &&
+\langle\chi|\delta_{Q_1 Q_2}{\cal E}_{\rm a}(R)|\chi\rangle 
+\langle\chi|\delta_{Q_1}{\cal E}_{\rm a}(R)|\delta_{Q_2}\chi\rangle
+\langle\chi|\delta_{Q_2}{\cal E}_{\rm a}(R)|\delta_{Q_1}\chi\rangle
+\ldots\nonumber\\ &=&
\langle Q_1 \, Q_2\rangle^{(0)} +\langle Q_1 \, Q_2\rangle^{(1)} +\ldots,
\end{eqnarray}
where $\delta_{Q_1 Q_2}{\cal E}_{\rm el}$ and $\delta_{Q_1 Q_2}{\cal E}_{\rm a}$
are corrections to corresponding energies due to electronic
operators $Q_1$ and $Q_2$, and
\begin{eqnarray}
\langle Q_1 \, Q_2\rangle^{(0)} &=&
\langle\chi|\langle\phiel|Q_1\,\frac{1}{({\cal E}_{\rm el}-H_{\rm el})'}\,Q_2|\phiel\rangle|\chi\rangle +
\nonumber \\ &&
+\langle\chi|\,\langle Q_1 \rangle_{\rm el}^{(0)}\,\frac{1}
{\bigl[ E_{\rm a} - H_{\rm n} - {\cal E}_{\rm a}(R) - {\cal E}_{\rm el}(R)\bigr]'}\,
\langle Q_2 \rangle_{\rm el}^{(0)}\,|\chi\rangle +{\rm c.c.}\,,
\label{57}
\end{eqnarray}
and
\begin{eqnarray}
\langle Q_1 \, Q_2\rangle^{(1)} &=&
\langle\chi|\,\langle\phiel|Q_1\,\frac{1}{({\cal E}_{\rm el}-H_{\rm el})'}\,
(\stackrel{\leftrightarrow}{H_{\rm n}}
-{\cal E}_{\rm a})\,\frac{1}{({\cal E}_{\rm el}-H_{\rm el})'}\,Q_2|
\phiel\rangle\,|\chi\rangle
\nonumber \\ &&
+\langle\chi|\,\langle\phiel|
\stackrel{\leftrightarrow}{H_{\rm n}}\,\frac{1}{({\cal E}_{\rm el}-H_{\rm el})'} 
\,(Q_1-\langle Q_1\rangle_{\rm el}^{(0)})
\,\frac{1}{({\cal E}_{\rm el}-H_{\rm el})'}\,Q_2|\phiel\rangle\,|\chi\rangle 
\nonumber \\ &&
+\langle\chi|\,\langle\phiel|Q_1\,\frac{1}{({\cal E}_{\rm el}-H_{\rm el})'}
\,(Q_2-\langle Q_2\rangle_{\rm el}^{(0)})\,
\frac{1}{({\cal E}_{\rm el}-H_{\rm el})'} \,\stackrel{\leftrightarrow}{H_{\rm n}}
|\phiel\rangle\,|\chi\rangle 
\nonumber \\ &&
+\langle\chi|\,\langle Q_1 \rangle_{\rm el}^{(1)}\,\frac{1}
{\bigl[ E_{\rm a} - H_{\rm n} - {\cal E}_{\rm a}(R) - {\cal E}_{\rm el}(R)\bigr]'}\,
\langle Q_2 \rangle_{\rm el}^{(0)}\,|\chi\rangle
\nonumber \\ &&
+\langle\chi|\,\langle Q_1 \rangle_{\rm el}^{(0)}\,\frac{1}
{\bigl[ E_{\rm a} - H_{\rm n} - {\cal E}_{\rm a}(R) - {\cal E}_{\rm el}(R)\bigr]'}\,
\langle Q_2 \rangle_{\rm el}^{(1)}\,|\chi\rangle
+ {\rm c.c.}\,.
\label{58}
\end{eqnarray}
In the case $\langle Q_1\rangle^{(0)}_{\rm el}=\langle Q_2\rangle^{(0)}_{\rm el} =
0$ above formulae coincide with those from the previous subsection.

\subsection{Matrix elements of Hermitian operators 
            with $R$-derivatives}
Consider an operator of the form 
$Q = \vec Q\cdot\vec P$
where $\vec P = -i\,\vec \nabla_R$, and $\vec Q$
is a Hermitian, electronic operator, such that $[P^i\,,\,Q^i] = 0$,
for example the electron-nucleus Breit interaction Eq. (\ref{65}).
Its matrix element between different rovibrational states is
of the form
\begin{eqnarray}
\langle\phi_{\rm el}\,\chi_{\rm f}|Q|\phi_{\rm el}\,\chi_{\rm i}\rangle &=&
\frac{i}{2}\,\bigl\langle\chi_{\rm f}\bigl|\Bigl[
\langle\vec\nabla_R\,\phi_{\rm el}|\vec Q|\phi_{\rm el}\rangle-
\langle\phi_{\rm el}|\vec Q|\vec\nabla_R\phi_{\rm el}\rangle\,
\Bigr]\bigr|\chi_{\rm i}\bigr\rangle\nonumber \\ &&
+\frac{i}{2}\,\Bigl[
\langle\vec\nabla\chi_{\rm f}|\langle \vec Q\rangle_{\rm el}|\chi_{\rm i}\rangle - 
\langle\chi_{\rm f}|\langle \vec Q\rangle_{\rm el}|\vec\nabla\chi_{\rm i}\rangle
\Bigr].
\label{59}
\end{eqnarray}
If $\vec Q$ is an imaginary operator, then the second term vanishes and
\begin{eqnarray}
\langle\phi_{\rm el}\,\chi_{\rm f}|Q|\phi_{\rm el}\,\chi_{\rm i}\rangle 
&=&\langle\chi_{\rm f}|\langle Q\rangle_{\rm el}|\chi_{\rm i}\rangle, \\
\langle Q\rangle_{\rm el} &=& 
-i\,\langle\phi_{\rm el}|\vec Q|\vec\nabla_R\,\phi_{\rm el}\rangle = 
i\,\langle\vec\nabla_R\,\phi_{\rm el}|\vec Q|\phi_{\rm el}\rangle.
\end{eqnarray}
If $\vec Q$ is a real operator, then the first term in Eq. (\ref{59}) vanishes and
\begin{eqnarray}
\langle\phi_{\rm el}\,\chi_{\rm f}|Q|\phi_{\rm el}\,\chi_{\rm i}\rangle 
&=&\langle\chi_{\rm f}|\langle \vec Q\rangle_{\rm el}\cdot\vec P|\chi_{\rm i}\rangle = 
\langle\chi_{\rm f}| \stackrel{\leftarrow}{P}\cdot\langle\vec Q\rangle_{\rm el}|\chi_{\rm i}\rangle.
\end{eqnarray}

\section{Relativistic recoil correction to rovibrational energies}
This is the first and the simplest application of
the nonadiabatic perturbation theory, the finite nuclear mass
correction to the relativistic energy. The total relativistic correction 
to the binding energy of a $\Sigma$ state, neglecting
interactions with nuclear spins and the higher order $O(m/M)^2$ terms, is
given by \cite{bs, magnetic}
\begin{eqnarray}
\delta H &=& \delta H_{\rm el} + \delta H_{\rm n},\\
\delta H_{\rm el} &=&\alpha^2\,\biggl[-\sum_a \frac{p_a^4}{8\,m^3}+
\sum_{a,X} \frac{Z_X\,\pi}{2\,m^2}\,\delta^3(r_{aX})
+\sum_{a>b}\frac{\pi}{m^2}\,\delta^3(r_{ab})
-\sum_{a>b}\,\frac{1}{2\,m^2}\,p_a^i\,
\biggl(\frac{\delta^{ij}}{r_{ab}} 
+ \frac{r_{ab}^i\,r_{ab}^j}{r_{ab}^3}\biggr)\,p_b^j\biggr]
\nonumber \\ \\
\delta H_{\rm n}  &=& \alpha^2\,\sum_{a,X}\frac{Z_X}{2\,m\,m_X}\,
p_a^i\biggl(\frac{\delta^{ij}}{r_{aX}}
+\frac{r_{aX}^i\,r_{aX}^j}{r_{aX}^3}\biggr)\,p_X^j.\label{65}
\end{eqnarray}
The relativistic correction $E_{\rm rel}$ to the energy, 
taking into account the transformation in Eq. (\ref{09}), is 
\begin{eqnarray}
E_{\rm rel} &=& \langle\phi|\delta H|\phi\rangle \\
         &=& \langle\chi|\langle\delta H\rangle_{\rm el}^{(0)}|\chi\rangle
                  +\langle\chi|\langle\delta H\rangle_{\rm
                    el}^{(1)}|\chi\rangle +O(m/M)^{3/2},
\end{eqnarray}
where
\begin{eqnarray}
\langle\delta H\rangle_{\rm el}^{(0)} &=& 
\langle\phi_{\rm el}|\delta H_{\rm el}|\phi_{\rm
  el}\rangle\equiv\langle\delta H_{\rm el}\rangle_{\rm el},\\
\langle\delta H\rangle_{\rm el}^{(1)} &=&
\langle\phi_{\rm el}|\delta H_{\rm n}|\phi_{\rm el}\rangle
-m\,\biggl(\frac{\varepsilon_A}{m_A}-\frac{\varepsilon_B}{m_B}\biggr)\,
\langle\phi_{\rm el}|[\sum_a\vec x_a\cdot\vec \nabla_R\,,\,\delta H_{\rm el}]|\phi_{\rm el}\rangle
\nonumber \\ &&
+2\,\langle\phi_{\rm el}|\delta H_{\rm el}\,
\frac{1}{({\cal E}_{\rm el}-H_{\rm el})'}
\stackrel{\leftrightarrow}{H_{\rm n}}|\phi_{\rm el}\rangle.
\label{69}
\end{eqnarray}
As it have already been noticed in \cite{jeziorsk},
relativistic recoil effects are of order $O(m/M)$ and can 
be expressed as a correction $\langle\delta H\rangle_{\rm el}^{(1)}$ 
to the BO energy ${\cal E}_{\rm el}(R)$. 

Since the wave function $\phi$ does not depend on $\vec R_G$, the momentum $\vec
P_G$ implicitly present in $\delta H_{\rm n}$ can be set to $0$. Moreover,
we chose $\vec R_G$ in dependence on the particular operator in $\delta H_{\rm el}$,
in such a way that the result of the transformation $\phi$ in Eq. (\ref{09}), namely the
second term in Eq. (\ref{69}) vanishes.
For example for the first, third and fourth term in $\delta  H_{\rm el}$
$\vec R_G$ is the nuclear mass center ($\varepsilon_A = m_A/(m_A+m_B),
\varepsilon_B = m_B/(m_A+m_B)$), for $\delta^{3}(r_{aX})$ $\vec R_G$ is
placed at the nucleus $X$ ($\varepsilon_X=1$), and for $\delta H_{\rm n}$
in the geometrical center ($\varepsilon_A=\varepsilon_B=1/2$). 
In this particular choice of $\vec R_G$, the derivative 
$\partial \delta H_{\rm el}/\partial R$  that comes from $H'_{\rm n}$
in Eq. (\ref{69}) also vanishes and the relativistic finite 
nuclear mass correction can be rewritten to the form
\begin{eqnarray}
\langle\delta H\rangle_{\rm el}^{(1)} &=&
\langle\phi_{\rm el}|\delta H_{\rm n}|\phi_{\rm el}\rangle
+2\,\langle\phi_{\rm el}|\delta H_{\rm el}\,
\frac{1}{({\cal E}_{\rm el}-H_{\rm el})'}
\biggl(H_{\rm n}''+\frac{\vec J_{\rm el}^2}{2\,m_{\rm n}\,R^2}\biggr)
|\phi_{\rm el}\rangle
\nonumber \\ &&
+\frac{1}{m_{\rm n}}\,
\langle\phi_{\rm el}|\frac{\partial V}{\partial R}\,
\frac{1}{({\cal E}_{\rm el}-H_{\rm el})'}\,
(\delta H_{\rm el} - \langle\delta H_{\rm el}\rangle_{\rm el})\,
\frac{1}{({\cal E}_{\rm el}-H_{\rm el})'^2}\,
\frac{\partial V}{\partial R}|\phi_{\rm el}\rangle
\nonumber \\ &&
+\frac{1}{m_{\rm n}}\,
\langle\phi_{\rm el}|\delta H_{\rm el}\,
\frac{1}{({\cal E}_{\rm el}-H_{\rm el})'}\,
\frac{\partial(V-{\cal E}_{\rm el})}{\partial R}\,
\frac{1}{({\cal E}_{\rm el}-H_{\rm el})'^2}\,
\frac{\partial V}{\partial R}|\phi_{\rm el}\rangle.
%\nonumber \\ &&
%+\frac{1}{m_{\rm n}}\,
%\langle\phi_{\rm el}|\vec\nabla_R(\delta H_{\rm el})\,
%\frac{1}{({\cal E}_{\rm el}-H_{\rm el})'^2}\,
%\vec\nabla_R(V)|\phi_{\rm el}\rangle
\end{eqnarray}
The expectation value of $\delta H_{\rm n}$ is calculated
according to Eq. (\ref{32}), namely if
\begin{equation}
\delta H_{\rm n} = \vec Q_1\cdot\vec\nabla_R+Q_2,
\end{equation} 
with $Q_i$ electronic operators and $\vec\nabla_R\,\vec Q_1 =0$, then
\begin{equation}
\langle\phi_{\rm el}|\delta H_{\rm n}|\phi_{\rm el}\rangle =
\langle\phi_{\rm el}|Q_2|\phi_{\rm el}\rangle+
\langle\phi_{\rm el}|\vec n\cdot\vec Q_1
\frac{1}{({\cal E}_{\rm el}-H_{\rm el})'}\frac{\partial V}{\partial R}|\phi_{\rm el}\rangle
+\frac{i}{R}\,\langle\phi_{\rm el}|\vec n\times\vec Q_1\cdot\vec J_{\rm n}|\phi_{\rm el}\rangle.
\end{equation} 
One may expect significant cancellations in the leading finite nuclear mass
correction $\langle\delta H\rangle_{\rm el}^{(1)}$ between the first and the
second term in Eq. (\ref{69}). For example for separate hydrogen
atoms in the  ground state, this correction vanishes. 
Therefore, the next order correction which is $O(m/M)^{3/2}$, 
may become relatively significant. It is due to,
for example, the second order in $H_{\rm n}$ nonadiabatic
correction given by  Eq. (\ref{37}), or by the orbit-orbit interaction between nuclei. 

\section{Electric properties}
We will study here the nonadiabatic corrections to the transition
dipole moment and the electric dipole static polarizability.
The direct nonadiabatic calculations have only been performed for simple
molecules like H$_2^+$ in \cite{polh2p1,polh2p2,polh2p3,polh2p4}, 
H$_2$ in \cite{adam1}, and LiH in \cite{adam2}.
There is a lot of literature on electric properties of molecules,
in what we call here, the adiabatic approximation. Let us mention
the extensive review of Bishop \cite{bishop} and earlier works of Brieger in
\cite{brieger1, brieger2}. We recover here their results for the electric
dipole polarizability, and present closed formulae for the nonadiabatic corrections.

The interaction of a neutral molecular system with the homogenous
electric field $\vec E$ is given by 
\begin{equation}
\delta H = -\vec D\cdot\vec E,
\end{equation}
where the electric dipole operator $\vec D$ is
\begin{equation}
\vec D = e\,\sum_a \vec x_a + e_A\,\vec x_A + e_B\,\vec x_B,
\end{equation}
and where $\vec x_A = \epsilon_B\,\vec R$ and $\vec x_B = -\epsilon_A\,\vec R$.
We note that for charged molecular systems, the electric dipole moment
has to be defined with respect to the mass center of the total molecule.

Let us fix the reference frame to the center of the nuclear charge
$\epsilon_A = e_A/(e_A+e_B)$, $\epsilon_B = e_B/(e_A+e_B)$,
then $\vec D = e\,\sum_a \vec x_a $
and this $\vec D$  is not affected by the unitary transformation in Eq. (\ref{09}).
We will show below that interaction of the molecule with 
the homogenous electric field, in spite of including leading finite nuclear
mass effects, can be effectively described by the Hamiltonian $H_{\rm eff}$ in the nuclear space
\begin{equation}
H_{\rm eff} = -\vec D_{\rm el}(\vec R)\cdot \vec E 
-\frac{\alpha^{ij}_{\rm el}(\vec R)}{2}\,E^i\,E^j\,,
\end{equation}
where $\vec D_{\rm el}(\vec R)$ is given in Eq. (\ref{78})
and $\alpha^{ij}_{\rm el}(\vec R)$ in Eq. (\ref{88}).
From the above equation, the transition dipole moment is
\begin{equation}
\vec D_{\rm fi} = \langle\chi_{\rm f}|\vec D_{\rm el}|\chi_{\rm i}\rangle,
\end{equation}
and one notes that the matrix elements $\vec D_{\rm fi}$ between 
the same nuclear states always vanish.
Similarly, the total electric dipole static polarizability is
\begin{equation}
\alpha^{ij} = \langle\chi|\alpha_{\rm el}^{ij}|\chi\rangle 
-2\,\langle\chi|D_{\rm el}^i\,\frac{1}{E_{\rm a} - H_{\rm n}
-{\cal E}_{\rm a}-{\cal E}_{\rm el}}\,D_{\rm el}^j|\chi\rangle,
\label{79}
\end{equation}
where the first term is due to the electron excitations and the second one
is due to the rovibrational excitations. For molecules, with so called
permanent electric dipole moment, the second term is dominating,
particularly significant contribution comes from 
intermediate nuclear states with the same vibrational number, but different $J$.

The electronic matrix elements $\vec D_{\rm el}$ and $\alpha_{\rm el}$
are obtained as follows.
According to Eqs. (\ref{35},\ref{40}) and Eqs. (\ref{29},\ref{30})
the electric dipole moment including the leading nonadiabatic effects is 
\begin{equation}
\vec D_{\rm el} = \vec D_{\rm el}^{(0)} + \vec D_{\rm el}^{(1)},\label{78}
\end{equation}
where
\begin{eqnarray}
\vec D_{\rm el}^{(0)} &=&\langle\phi_{\rm el}|\vec D|\phi_{\rm el}\rangle,\\
\vec D_{\rm el}^{(1)} &=&2\,\langle\phi_{\rm el}|\vec D\,
\frac{1}{({\cal E}_{\rm el}-H_{\rm el})'}
\stackrel{\leftrightarrow}{H_{\rm n}}|\phi_{\rm el}\rangle.
\end{eqnarray}
For $\Sigma$ electronic states the nonadiabatic correction
can be rewritten to the form
\begin{eqnarray} 
\vec D_{\rm el}^{(1)} &=& 2\,\langle\phi_{\rm el}|\vec D
\frac{1}{({\cal E}_{\rm el}-H_{\rm el})'}
\biggl(H''_{\rm n}+\frac{\vec J_{\rm el}^2}{2\,m_{\rm n}\,R^2}
\biggr)|\phi_{\rm el}\rangle
+\frac{i}{m_{\rm n}\,R^2}\,\langle\phi_{\rm el}|\vec D\times
\frac{1}{({\cal E}_{\rm el}-H_{\rm el})'}\vec J_{\rm el}|\phi_{\rm el}\rangle
\nonumber \\ &&
+\frac{1}{m_{\rm n}}\,
\langle\phi_{\rm el}|\vec D
\frac{1}{({\cal E}_{\rm el}-H_{\rm el})'}\,
\frac{\partial(V-{\cal E}_{\rm el})}{\partial R}\,
\frac{1}{({\cal E}_{\rm el}-H_{\rm el})'^2}\,
\frac{\partial V}{\partial R}\,
|\phi_{\rm el}\rangle
\nonumber \\ &&
+\frac{1}{m_{\rm n}}\,
\langle\phi_{\rm el}|\frac{\partial V}{\partial R}\,
\frac{1}{({\cal E}_{\rm el}-H_{\rm el})'}\,
(\vec D-\vec D_{\rm el}^{(0)})\,
\frac{1}{({\cal E}_{\rm el}-H_{\rm el})'^2}\,
\frac{\partial V}{\partial R}\,
|\phi_{\rm el}\rangle.
\end{eqnarray}
This result is in agreement with the previously obtained 
much simpler formula for the HD molecule \cite{hd}, 
where only the last term in $H_{\rm n}$ contributes in the above equation,
due to the inversion symmetry of the ground electronic state.
We observe that in spite of including first order nonadiabatic corrections,
the transition dipole moment $\vec D_{\rm fi}$ can be represented
in terms of the matrix element of the electronic dipole moment 
$\vec D_{\rm el}$ evaluated with nuclear functions $\chi_{\rm f}^*$
and $\chi_{\rm i}$, see Eq. (\ref{39}). A similar result holds for the
electric dipole static polarizability
\begin{equation}
\alpha_{\rm el}^{ij} = \alpha^{(0)\,ij}_{\rm el} + \alpha^{(1)\,ij}_{\rm el},\label{88}
\end{equation}
where in the BO approximations it is
\begin{equation}
\alpha^{(0)\,ij}_{\rm el} = -2\,
\langle\phi_{\rm el}|D^i\,\frac{1}{({\cal E}_{\rm el}-H_{\rm el})'}
\,D^j|\phi_{\rm el}\rangle,
\label{89}
\end{equation} 
and the nonadiabatic correction, using Eq. (\ref{58}) is
\begin{eqnarray}
\alpha^{(1)\,ij}_{\rm el} &=& -2\,\biggl[
\langle\phi_{\rm el}|D^i\,\frac{1}{({\cal E}_{\rm el}-H_{\rm el})'}
(\stackrel{\leftrightarrow}{H_{\rm n}}-{\cal E}_{\rm a})
\,\frac{1}{({\cal E}_{\rm el}-H_{\rm el})'}
\,D^j|\phi_{\rm el}\rangle\nonumber \\ &&
+\langle\phi_{\rm el}|
\stackrel{\leftrightarrow}{H_{\rm n}}\,\frac{1}{({\cal E}_{\rm el}-H_{\rm el})'}
(D^i-D^{(0)\,i}_{\rm el})
\,\frac{1}{({\cal E}_{\rm el}-H_{\rm el})'}
\,D^j|\phi_{\rm el}\rangle\nonumber \\ &&
+\langle\phi_{\rm el}|D^i\,\frac{1}{({\cal E}_{\rm el}-H_{\rm el})'}
(D^j-D^{(0)\,j}_{\rm el})
\,\frac{1}{({\cal E}_{\rm el}-H_{\rm el})'}
\,\stackrel{\leftrightarrow}{H_{\rm n}}|\phi_{\rm el}\rangle\biggr].
\label{90}
\end{eqnarray}
The explicit formula after taking the derivative with respect to $R$
is little bit too long to be written here,
but can easily be obtained with the help of Eqs. (\ref{29},\ref{30},\ref{31}).
Eq. (\ref{79}) in the adiabatic approximation, namely 
$\alpha_{\rm el}\rightarrow \alpha_{\rm el}^{(0)}$, 
$\vec D_{\rm el}\rightarrow\vec D_{\rm el}^{(0)}$ was already obtained in the
literature \cite{brieger1, brieger2, bishop}. Results for nonadiabatic
corrections when $\alpha_{\rm el}\rightarrow \alpha_{\rm el}^{(1)}$, 
$\vec D_{\rm el}\rightarrow\vec D_{\rm el}^{(1)}$ are, in our opinion new
and have not been discussed in the literature.

\section{Magnetic properties}
We consider here nonadiabatic corrections to magnetic properties 
of a diatomic molecule. These properties within the BO approximation
are reviewed in details by Flygare in \cite{flygare}. Here, we demonstrate
that the nonadiabatic corrections  can be implemented in the effective nuclear
Hamiltonian, similarly to that in the BO approximation, namely
\begin{equation}
H_{\rm eff} = -\eta(R)\,\vec I\cdot\vec J -\gamma_I\,\vec I(1-\hat\sigma_{\rm el}(\vec R))\vec B
-\gamma_J(R)\,\vec J\cdot\vec B - \frac{1}{2}\,\vec B\hat\chi(\vec R)\vec B,
\label{91}
\end{equation}
where it is understood that the electronic part of the angular momentum
operator $\vec J_{\rm el}$ Eq. (\ref{17}) vanishes on $\chi$,
$\eta$ is the spin-rotation constant, 
$\mu_I=I\,\gamma_I$ is the nuclear magnetic moment, 
$\mu_J=J\,\gamma_j$ is the orbital magnetic 
moment, $\sigma_{\rm el}$ is the $R$-dependent shielding constant, $\hat\chi$ is the magnetic
susceptibility, and $\vec B$ is the magnetic field.
The Hamiltonian in Eq. (\ref{91}) should in principle
involve also quadrupolar interaction
$(I^i\,I^j)^{(2)}\,(J^i\,J^j)^{(2)}$ or $(I^i\,I^j)^{(2)}\,J^i\,B^j$.
The first term comes from the quadrupole moment of nucleus or 
tensor interaction between nuclear magnetic moments, 
while the second term has not been investigated in the literature so far.
Both terms will not be considered in this work.
If two different nuclei are involved, each one has its own spin and magnetic
moments, then interaction between them should be included,
but we do not consider it either. In the following we 
rederive in a simple way known results for rotational magnetic moment $\mu_{\rm J}$ and 
spin rotation constant $\eta$ and obtain nonadiabatic corrections
to the shielding constant. These corrections, up to our knowledge,
have not yet been investigated in the literature \cite{hunt}. 

\subsection{Nonrelativistic Hamiltonian in the external magnetic field}
In order to obtain finite nuclear mass corrections we start with the
Hamiltonian of a molecular system in the homogenous magnetic field
\begin{equation}
H_0 = \sum_\beta\,\frac{\vec\pi_\beta^2}{2\,m_\beta} + V,
\end{equation}
where $\beta$ enumerates electrons and nuclei.
We assume the coordinate system as defined in Eqs. (\ref{02}-\ref{04}): 
$\vec x_\beta = \vec r_\beta-\vec R_G$  and perform
the following unitary transformation
\begin{eqnarray}
\tilde H_0 &=& e^{-i\,\varphi}\,H_0\,e^{i\,\varphi}+\partial_t\varphi,\\
\varphi &=& \sum_\beta e_\beta\int_0^1 du\,\vec x_\beta\cdot\vec A(\vec R_G+u\,\vec x_\beta) 
\nonumber \\
     &=& \sum_\beta e_\beta\bigl[ x_\beta^i\,A^i + x_\beta^i\,x_\beta^j\,A^i_{,j}/2\bigr], \label{94}\\
\pi_\beta^j &=& p_\beta^j -e_\beta\,A^j(\vec r_\beta)\nonumber \\
          &=& p_\beta^j -e_\beta\,\bigl[A^j+x_\beta^i\,A^j_{,i}\bigr],
\end{eqnarray}
where $\vec A \equiv \vec A(\vec R_G)$. The result of this 
transformation on the canonical momentum is
\begin{eqnarray}
e^{-i\,\varphi}\,\pi^i_a\,e^{i\,\varphi} &=& p_a^i+\frac{e_a}{2}\,(\vec x_a\times\vec B)^j ,\\
e^{-i\,\varphi}\,\pi^i_A\,e^{i\,\varphi} &=& p_A^i+\frac{e_A}{2}\,(\vec x_A\times\vec B)^j+
\frac{\epsilon_A}{2}\,(\vec D\times\vec B)^j,\\
e^{-i\,\varphi}\,\pi^i_B\,e^{i\,\varphi} &=& p_B^i+\frac{e_B}{2}\,(\vec x_B\times\vec B)^j+
\frac{\epsilon_B}{2}\,(\vec D\times\vec B)^j,
\end{eqnarray}
where $\vec D$ is the total electric dipole moment 
$\vec D = \sum_\beta e_\beta\,\vec x_\beta$
and we assumed that $\sum_\beta e_\beta = 0$, so the molecule is neutral.
In consequence, the transformed Hamiltonian does not depend on $\vec R_G$,
thus we are allowed to set $\vec P_G = 0$. In order to further simplify,
we perform the next transformation as defined in Eq. (\ref{09}), 
neglect $O(m/M)^2$ terms, and the transformed Hamiltonian takes the form
\begin{eqnarray}
H_0 &=& H_{\rm el} + H_{\rm n} + H_{\mu} + H_{\chi},\label{99}\\
H_{\mu} &=& -\sum_a\,\frac{e}{2\,m}\,\vec x_a\times\vec q_a\cdot\vec B +\delta H_\mu,
\label{100}\\
\delta H_\mu  &=& 
-\biggl(\frac{\epsilon_A}{m_A}-\frac{\epsilon_B}{m_B}\biggr)\,\vec D\times\vec P\cdot\vec B
-\frac{1}{2}\,\biggl(\frac{1}{m_A}+\frac{1}{m_B}\biggr)\,
(e_A\,\epsilon_B^2+e_B\,\epsilon_A^2)\,\vec R\times\vec P\cdot\vec B
\nonumber \\ &&
+\frac{1}{2}\,\biggl(\frac{\epsilon_A^2}{m_A}+\frac{\epsilon_B^2}{m_B}\biggr)\,
\vec D\times\sum_a\vec q_a\cdot\vec B
+\frac{\epsilon_A\,\epsilon_B}{2}\,\biggl(\frac{e_A}{m_A}-\frac{e_B}{m_B}\biggr)\,
\vec R\times\sum_a \vec q_a\cdot\vec B,
\label{101}\\
H_{\chi} &=& \sum_a\,\frac{e^2}{8\,m}\,(\vec x_a\times\vec B)^2 +\delta H_\chi,\\
\delta H_\chi  &=& 
\frac{1}{8\,m_A}\,(e_A\,\vec x_A\times\vec B+\epsilon_A\,\vec D\times\vec B)^2
+\frac{1}{8\,m_B}\,(e_B\,\vec x_B\times\vec B+\epsilon_B\,\vec D\times\vec B)^2.
\end{eqnarray}
The center of the reference frame $\vec R_G$ is placed arbitrarily on the symmetry
axis. This freedom will be used below to simplify formulae for nonadiabatic
corrections.

\subsection{Rotational magnetic moment}
The rotational magnetic moment has been first investigated by Wick in \cite{wick} for H$_2$,
later by Ramsey in who extended Wick's result to
arbitrary masses of nuclei and performed first measurements in Ref. \cite{ram1}
and improved measurements with Harrick in Ref. \cite{ram5}.
The rotational magnetic moments results from coupling of
the molecular rotation to the magnetic field.
The expectation value on the $\Sigma$ state of the first term in Eq. (\ref{100})
$e/(2\,m)\,\vec J_{\rm el}\cdot\vec B$ vanishes, 
thus the leading coupling comes from nonadiabatic corrections
to the matrix element as given by Eq. (\ref{43}) and from $\delta H_{\mu}$, namely 
\begin{equation}
\gamma_J = 
\biggl(\frac{\epsilon_A}{m_A}-\frac{\epsilon_B}{m_B}\biggr)\,
\frac{\vec D_{\rm el}\cdot \vec R}{R^2}
+\frac{(e_A\,\epsilon_B^2+e_B\,\epsilon_A^2)}{2\,m_{\rm n}}
-\frac{e}{2\,m}\,\frac{1}{m_{\rm n}\,R^2}\,
\langle\phi_{\rm el}|\vec J_{\rm el}\,\frac{1}{({\cal E}_{\rm el}-H_{\rm el})'} 
\,\vec J_{\rm el}|\phi_{\rm el}\rangle.\label{104}
\end{equation}
In the center of nuclear mass frame $\epsilon_A = m_A/(m_A+m_B)$, 
$\epsilon_B = m_B/(m_A+m_B)$ the rotational magnetic moment 
in units of the nuclear magneton $\mu_I$ becomes
\begin{equation}
\frac{\gamma_J}{\mu_I} = \frac{\mu_{\rm J}}{J\,\mu_I} =
m_{\rm n}\,m_p\,\biggl(\frac{Z_A}{m_A^2}+\frac{Z_B}{m_B^2}\biggr)
+\frac{m_p}{m_{\rm n}\,m\,R^2}\,
\langle\phi_{\rm el}|J_{\rm el}^i\,\frac{1}{({\cal E}_{\rm el}-H_{\rm el})'} 
\,J_{\rm el}^i|\phi_{\rm el}\rangle,
\end{equation}
where $m_p$ is the proton mass, in agreement with the result from Ref. \cite{ram1}
[Eq. (4) with assuming $Z_A=Z_B=1$]. 
One observes that rotational magnetic moment in Eq. (\ref{104}) 
does not depend on the choice of reference
frame, but depends on the distance $R$ between nuclei.
For large $R$ it vanishes at least with $R^{-6}$.
The dependence of $\gamma_J$ on $R$ leads to the appearance of the magnetic
dipole transition in $H_2$, between states of the same angular momentum,
but of different vibrational number. These transitions, up to our knowledge,
have not yet been investigated in the literature and may play a role
in the astrophysical environment, however their importance should be verified
by explicit calculations. 

As was first noted in Ref. \cite{ram5}  the rotational magnetic moment is 
related to the paramagnetic part of magnetic susceptibility $\chi$.
In the BO approximation $\chi_{\rm el}$ is given by
\begin{equation}
\chi_{\rm el}^{ij} = 
-\frac{e^2}{4\,m}\,\sum_a\,\langle\phi_{\rm el}|\vec x_a^2\delta^{ij}-x_a^i\,x_a^j|\phi_{\rm el}\rangle
-\frac{e^2}{2\,m^2}\,\langle\phi_{\rm el}|J_{\rm el}^i
\,\frac{1}{({\cal E}_{\rm el}-H_{\rm el})'}\,
J_{\rm el}^j|\phi_{\rm el}\rangle.
\end{equation}
For $\Sigma$ states it can be simplified to the form
\begin{eqnarray}
\chi_{\rm el}^{ij} &=& 
-\frac{e^2}{8\,m}\,\sum_a\Bigl[2\,\langle\phi_{\rm el}| 
\vec x_a^2 - (\vec n\cdot\vec x_a)^2|\phi_{\rm el}\rangle\,n^i\,n^j
+\langle\phi_{\rm el}| 
\vec x_a^2 + (\vec n\cdot\vec x_a)^2|\phi_{\rm el}\rangle\,(\delta^{ij}-n^i\,n^j)
\nonumber \\ &&
-\frac{e^2}{4\,m^2}\,\langle\phi_{\rm el}|\vec J_{\rm el}
\,\frac{1}{({\cal E}_{\rm el}-H_{\rm el})'}\,
\vec J_{\rm el}|\phi_{\rm el}\rangle\,\bigl(\delta^{ij}-n^i\,n^j\bigr),
\label{107}
\end{eqnarray}
The last terms in both Eqs. (\ref{104}) and (\ref{107}),
which are the second order matrix elements, are similar
while the first terms in both equations are simple
to evaluate. This allows one to express the difficult to measure
magnetic susceptibility in terms of the rotational magnetic moment.
This relation however, works only in the BO approximation, as
nonadiabatic corrections will be different.
As noticed by Authors of \cite{herman}, this second order matrix element 
with $\vec J_{\rm el}$ operator, is present also
in the nonadiabatic equation for the nuclear function $\chi$ as the $W_\perp$
function (in the notation from our previous work \cite{nonad}).
 
\subsection{Spin-rotation Hamiltonian}
The general spin-orbit Hamiltonian for arbitrary nuclei,
including the external magnetic field is $(\hbar=c=1)$ \cite{bs,magnetic,gyro}
\begin{equation}
\delta H = \sum_{\alpha,\beta}\frac{e_\alpha\,e_\beta}{4\,\pi}\,\frac{1}{2\,r_{\alpha\beta}^3}\,
\biggl[\frac{g_\alpha}{m_\alpha\,m_\beta}\,\vec s_\alpha\cdot\vec r_{\alpha\beta}\times\vec \pi_\beta-
\frac{(g_\alpha-1)}{m_\alpha^2}\,\vec s_\alpha\cdot\vec r_{\alpha\beta}\times \vec\pi_\alpha \biggr],
\label{108}
\end{equation}
where summation over $\alpha,\,\beta$ goes over electrons and nuclei. 
In particular, the coupling of the nuclear spin $\vec I = \vec s_A$ 
to the rotation and to the magnetic field using Eq. (\ref{108}) is
\begin{eqnarray}
\delta H &=& \sum_{b}\frac{e_A\,e}{4\,\pi}\,\frac{\vec I}{2\,r_{A b}^3}\,
\biggl[\frac{g_A}{m_A\,m}\,\vec r_{A b}\times\vec \pi_b-
       \frac{(g_A-1)}{m_A^2}\,\vec r_{A b}\times \vec\pi_A \biggr]
\nonumber \\ &&
+\frac{e_A\,e_B}{4\,\pi}\,\frac{\vec I}{2\,r_{A B}^3}\,
\biggl[\frac{g_A}{m_A\,m_B}\,\vec r_{A B}\times\vec \pi_B-
       \frac{(g_A-1)}{m_A^2}\,\vec r_{A B}\times \vec\pi_A \biggr].
\label{109}
\end{eqnarray} 
For convenience we chose the reference frame centered at
the considered nucleus $\vec R_G = \vec R_A$, so $\epsilon_A = 1$,$\epsilon_B = 0$, 
perform unitary transformations in Eqs. (\ref{94},\ref{09}), and obtain
\begin{eqnarray}
\delta H &=& -\sum_{b}\frac{e_A\,e}{4\,\pi}\,\frac{\vec I}{2\,m_A}\times\frac{\vec x_b}{x_b^3}\cdot
\biggl[\frac{g_A}{m}
\biggl(\vec q_b+\frac{m}{m_A}\vec P +\frac{e}{2}\,\vec x_b\times\vec B\biggr)-
\frac{(g_A-1)}{m_A}\biggl(\vec P-\vec q_{\rm el}+\frac{\vec D}{2}\times\vec B\biggr) \biggr]
\nonumber \\ &&
-\frac{e_A\,e_B}{4\,\pi}\,\frac{\vec I}{2\,m_A}\times\frac{\vec R}{R^3}\cdot
\biggl[\frac{g_A}{m_B}\,\biggl(\vec P+\frac{e_B}{2}\,\vec R\times\vec B\biggr)+
       \frac{(g_A-1)}{m_A}\,\biggl(\vec P-\vec q_{\rm el}+\frac{\vec D}{2}\times\vec B\biggr) \biggr],
\label{110}
\end{eqnarray} 
where $\vec q_{\rm el} = \sum_a\vec q_a$, $\vec x_{\rm el} = \sum_a\vec x_a$,
and the electric dipole operator is $\vec D=e\,\vec x_{\rm el}-e_B\,\vec R$. 
This Hamiltonian will be used in
next sections to rederive the known formulae for the spin-rotation and the
shielding constants, and to obtain a closed expression for 
the nonadiabatic corrections to the magnetic shielding, which contribute at the level of 
$m_{\rm e}/m_{\rm n}$ what for H$_2$ is about $10^{-3}$. 
 
\subsection{Spin-rotation constant}
The theory of the spin-rotation interaction was introduced by Wick \cite{wick},
Ramsey \cite{ram1}, and Foley \cite{foley},
and further developed by Ramsey \cite{ram2, ram3}, by Frosch and Foley
\cite{frosch} and again by Ramsey \cite{ram4}.
These theoretical results were not in agreement for a long time, until
Reid and Chu in \cite{reid} found a complete set of corrections.
Here we rederive their result on the basis of Eq. (\ref{110}).
 
The spin-rotation interactions results from $\delta H$ above with $\vec B =
0$. For considered $\Sigma$ states terms with $\vec q_{\rm el}$ vanish and
$\delta H$ takes the form
\begin{eqnarray}
\delta H &=& \vec Q_1\cdot\vec I + \vec Q_2\times\vec P\cdot\vec I,\\
\vec Q_1
&=&-\sum_{b}\frac{e_A\,e}{4\,\pi}\,\frac{g_A}{2\,m\,m_A}\,\frac{\vec x_b\times \vec q_b}{x_b^3},\\
\vec Q_2 &=& -\sum_{b}\frac{e_A\,e}{4\,\pi}\,\frac{1}{2\,m_A^2}\,\frac{\vec x_b}{x_b^3}
-\frac{e_A\,e_B}{4\,\pi}\,\frac{1}{2\,m_A}\,
\biggl[\frac{g_A}{m_B}+\frac{(g_A-1)}{m_A} \biggr]\,\frac{\vec R}{R^3}.
\end{eqnarray} 
The expectation value of $\langle\phi_{\rm el}|\vec Q_1|\phi_{\rm el}\rangle$
vanishes, and $\vec Q_1$ operator contributes only through nonadiabatic matrix elements
Eq. (\ref{43}), so the total spin-rotation constant is given by
\begin{equation}
-\eta\,\vec I\cdot\vec J = -\frac{\vec I\cdot\vec J}{m_{\rm n}\,R^2}\,
\langle\phi_{\rm el}|\vec J_{\rm el}\,\frac{1}{({\cal E}_{\rm el}-H_{\rm el})'}\,\vec
Q_1|\phi_{\rm el}\rangle + 
\langle\phi_{\rm el}|\vec Q_2|\phi_{\rm el}\rangle\times\vec P\cdot\vec I.
\end{equation}
The expectation value of the first term in $\vec Q_2$ can be expressed in
terms of derivative of BO energy, namely
\begin{eqnarray}
\langle\phi_{\rm el}|\sum_{b}\frac{e_A\,e}{4\,\pi}\,\frac{\vec x_b}{x_b^3}|\phi_{\rm el}\rangle
&=&
\vec n\,\biggl(\frac{\partial{\cal E}_{\rm el}}{\partial R}+\frac{e_A\,e_B}{4\,\pi}\,\frac{1}{R^2}\biggr),
\end{eqnarray}
and thus $\eta$ in atomic units becomes $[e_X =-Z_X\,e, \alpha=e^2/(4\,\pi)]$ 
\begin{eqnarray}
\eta &=&\alpha^2\biggl[\frac{1}{R^2}\,\frac{Z_A\,g_A}{2\,m_{\rm n}\,m_A}\,
\langle\phi_{\rm el}|\sum_a\vec x_a\times\vec q_a\,\frac{1}{({\cal E}_{\rm el}-H_{\rm el})'}
\sum_{b} \frac{\vec x_b\times \vec q_b}{x_b^3}|\phi_{\rm el}\rangle \nonumber \\ &&
+ \frac{1}{R}\,\frac{1}{2\,m_A^2}\,\frac{\partial{\cal E}_{\rm el}}{\partial R}
+\frac{1}{R^3}\frac{Z_A\,Z_B\,g_A}{2\,m_A\,m_{\rm n}}\biggr],
\end{eqnarray}
in agreement with Ref. \cite{reid}, their $C$ is related to our $\eta$ by $C=2\,\pi\,\eta$. 
Since, there is cancellation between the first and the third term, 
$\eta$ vanishes at least as $R^{-6}$ for large values of $R$.

\subsection{Magnetic shielding constant}
The shielding of the external magnetic field due to atomic electrons,
was first considered by Ramsey in
\cite{ram2} with the help of the Breit-Pauli Hamiltonian with the external magnetic
field. Here we rederive his result
and in addition to it, obtain nonadiabatic corrections. 
A similar calculations for atoms have recently been performed in
\cite{magnetic,rudzins}.

The Hamiltonian of a molecule
in the magnetic field including the nuclear spin, but neglecting that of electrons,
is a sum of $H_0$ in Eq. (\ref{99}) and $\delta H$ in Eq. (\ref{109}).
In the BO approximation $\hat\sigma^{(0)}_{\rm el}$ is a sum of diamagnetic and
paramagnetic parts. In atomic units, they are correspondingly
\begin{eqnarray}
\vec I\,\hat\sigma^{(0)}_{\rm el}\vec B &=& \alpha^2\,\biggl[-\frac{1}{2}\,
\langle\phi_{\rm el}|\sum_{b}\biggl(\vec I\times\frac{\vec x_b}{x_b^3}\biggr)\cdot
\bigl(\vec x_b\times\vec B\bigr) |\phi_{\rm el}\rangle
\nonumber \\ &&
+\langle\phi_{\rm el}|\sum_a\vec x_a\times\vec q_a\cdot \vec B\,
\frac{1}{({\cal E}-H_{\rm el})'}\,
\sum_{b}\frac{\vec x_b\times\vec q_b}{x_b^3}\cdot\vec I|\phi_{\rm el}\rangle\biggr].
\label{117}
\end{eqnarray}
Nonadiabatic corrections $\hat\sigma^{(1)}_{\rm el}$, namely all corrections which are linear in the
electron-nuclear mass ratio come from several sources, and we split them into four parts
\begin{equation}
\hat\sigma^{(1)}_{\rm el} = \hat\sigma^{(1)}_{\rm n} + \hat\sigma^{(1)}_{\rm d} +
\hat\sigma^{(1)}_{\rm s} + \hat\sigma^{(1)}_{\rm l}. 
\end{equation}
$\hat\sigma^{(1)}_{\rm n}$ is the correction due to $H_{\rm n}$ to the matrix
elements in Eq. (\ref{117}), namely using Eq. (\ref{49}) one obtains
\begin{eqnarray}
\vec I\,\hat\sigma^{(1)}_{\rm n}\vec B &=&\alpha^2\,\biggl[
-\langle\phi_{\rm el}|\sum_{b}\biggl(\vec I\times\frac{\vec x_b}{x_b^3}\biggr)\cdot
\bigl(\vec x_b\times\vec B\bigr)\,\frac{1}{({\cal E}-H_{\rm el})'}
\,\stackrel{\leftrightarrow}{H_{\rm n}} |\phi_{\rm el}\rangle
\nonumber \\ &&
+\langle\phi_{\rm el}|\sum_a\vec x_a\times\vec q_a\cdot \vec B\,
\frac{1}{({\cal E}-H_{\rm el})'}
\,(\stackrel{\leftrightarrow}{H_{\rm n}}-{\cal E}_{\rm a})
\,\frac{1}{({\cal E}-H_{\rm el})'}\,
\sum_{b}\frac{\vec x_b\times\vec q_b}{x_b^3}\cdot\vec I|\phi_{\rm el}\rangle
\nonumber \\ &&
+\langle\phi_{\rm el}|
\stackrel{\leftrightarrow}{H_{\rm n}}
\,\frac{1}{({\cal E}-H_{\rm el})'}\,
\sum_a\vec x_a\times\vec q_a\cdot \vec B\,
\frac{1}{({\cal E}-H_{\rm el})'}
\sum_{b}\frac{\vec x_b\times\vec q_b}{x_b^3}\cdot\vec I|\phi_{\rm el}\rangle
\nonumber \\ &&
+\langle\phi_{\rm el}|\sum_a\vec x_a\times\vec q_a\cdot \vec B
\,\frac{1}{({\cal E}-H_{\rm el})'}\,
\sum_{b}\frac{\vec x_b\times\vec q_b}{x_b^3}\cdot\vec I\,
\frac{1}{({\cal E}-H_{\rm el})'}
\,\stackrel{\leftrightarrow}{H_{\rm n}}|\phi_{\rm el}\rangle\biggr].
\end{eqnarray}
$\hat\sigma^{(1)}_{\rm d}$ is a correction to the diamagnetic part
due to the direct coupling of the nuclear spin to the
magnetic field in $\delta H$, Eq. (\ref{110})
\begin{eqnarray}
\vec I\,\hat\sigma^{(1)}_{\rm d}\vec B &=&
\frac{\alpha^2}{2\,g_A}\,\langle\phi_{\rm el}\biggl|
\frac{(g_A-1)}{m_A}\,\sum_{b}\biggl(\vec I\times\frac{\vec x_b}{x_b^3}\biggr)\cdot
(\vec x_{\rm el}+Z_B\,\vec R)\times\vec B 
\nonumber \\ &&
-\biggl(\vec I\times\frac{\vec R}{R^3}\biggr)\cdot
\biggl[\frac{Z_B^2}{m_B}\,\vec R-
       \frac{Z_B\,(g_A-1)}{m_A}\,(\vec x_{\rm el}+Z_B\,\vec R) \biggr]\times\vec B
\biggr|\phi_{\rm el}\rangle. 
\end{eqnarray}
$\hat\sigma^{(1)}_{\rm s}$ is a correction to paramagnetic part
due to spin-rotation interaction  in $\delta H$ Eq. (\ref{110})
\begin{eqnarray}
\vec I\,\hat\sigma^{(1)}_{\rm s}\vec B &=&
\frac{\alpha^2}{g_A}\,\langle\phi_{\rm el}|\sum_a\vec x_a\times\vec q_a\cdot\vec B\,
\frac{1}{({\cal E}-H_{\rm el})'}
\biggl\{
\frac{1}{m_A}\,\sum_{b}\vec I\cdot\frac{\vec x_b}{x_b^3}\times
\Bigl[\vec P+(g_A-1)\,\vec q_{\rm el} \Bigr]
\nonumber \\ &&
-Z_B\,\vec I\cdot\frac{\vec R}{R^3}\times
\biggl[\frac{g_A}{m_B}\,\vec P+
       \frac{(g_A-1)}{m_A}\,\bigl(\vec P-\vec q_{\rm el}\bigr)
       \biggr]\biggr\}|\phi_{\rm el}\rangle.
\end{eqnarray}
The derivative over nuclear coordinates $\vec P$ in the above and below,
can act on the right or on the left, 
since these matrix elements does not depend on this.
Finally, $\hat\sigma^{(1)}_{\rm l}$ is a correction to the paramagnetic
part due to $\delta H_\mu$ in Eq. (\ref{101}),
\begin{eqnarray}
\vec I\,\hat\sigma^{(1)}_{\rm l}\vec B &=&
\alpha^2\,\langle\phi_{\rm el}|
\biggl\{ 
\frac{1}{m_A}\,\bigl(\vec x_{\rm el}+Z_B\,\vec R\bigr)\times
\bigl(2\,\vec P-\vec q_{\rm el}\bigr)
-\frac{Z_B}{m_{\rm n}}\,\vec R\times\vec P
\biggr\}\cdot\vec B
\nonumber \\ &&
\frac{1}{({\cal E}-H_{\rm el})'}\,
\sum_b\frac{\vec x_b\times\vec q_b}{x_b^3}\cdot\vec I|\phi_{\rm el}\rangle.
\end{eqnarray}
The total magnetic shielding $\hat\sigma$
is obtained by averaging with the nuclear wave function
\begin{equation}
\sigma^{ij} = \langle\chi|
\sigma^{(0)ij}_{\rm el} +\sigma^{(1)ij}_{\rm el}|\chi\rangle,
\end{equation}
and one notes, that the orbital magnetic moment with the spin-rotation
coupling gives additional contribution to the shielding constant
\begin{equation}
\delta\sigma^{ij} =
\frac{2}{\gamma_I}\,\langle\chi|\eta\,J^i\,
\frac{1}{[E_{\rm a}-H_{\rm n}-{\cal E}_{\rm a}-{\cal E}_{\rm
      el}]'}\,\gamma_L\,J^j|\chi\rangle,
\end{equation}
which however is negligible.

Let us present in more details the averaged shielding $\sigma=\sigma^{ii}/3$ 
in the case of H$_2$ and isotopomers $Z_A=Z_B=1$,
\begin{equation}
\sigma = \langle\chi|
\sigma^{(0)}_{\rm el} +\sigma^{(1)}_{\rm el}|\chi\rangle, 
\end{equation}
and
\begin{eqnarray}
\sigma^{(0)}_{\rm el} &=& \frac{\alpha^2}{3}\,\biggl[
\langle\phi_{\rm el}|\sum_{b}\frac{1}{x_b} |\phi_{\rm el}\rangle
+\langle\phi_{\rm el}|\sum_a\vec x_a\times\vec q_a
\frac{1}{({\cal E}-H_{\rm el})'}\,
\sum_{b}\frac{\vec x_b\times\vec q_b}{x_b^3}|\phi_{\rm el}\rangle\biggr],\\
\sigma^{(1)}_{\rm n} &=&\frac{\alpha^2}{3}\,\biggl[
2\,\langle\phi_{\rm el}|\sum_{b}\frac{1}{x_b}\,\frac{1}{({\cal E}-H_{\rm el})'}
\,\stackrel{\leftrightarrow}{H_{\rm n}} |\phi_{\rm el}\rangle
\nonumber \\ &&
+\langle\phi_{\rm el}|\sum_a\vec x_a\times\vec q_a
\frac{1}{({\cal E}-H_{\rm el})'}
\,(\stackrel{\leftrightarrow}{H_{\rm n}}-{\cal E}_{\rm a})
\,\frac{1}{({\cal E}-H_{\rm el})'}\,
\sum_{b}\frac{\vec x_b\times\vec q_b}{x_b^3}|\phi_{\rm el}\rangle
\nonumber \\ &&
+\langle\phi_{\rm el}|
\stackrel{\leftrightarrow}{H_{\rm n}}
\,\frac{1}{({\cal E}-H_{\rm el})'}\,
\sum_a\vec x_a\times\vec q_a
\frac{1}{({\cal E}-H_{\rm el})'}
\sum_{b}\frac{\vec x_b\times\vec q_b}{x_b^3}|\phi_{\rm el}\rangle
\nonumber \\ &&
+\langle\phi_{\rm el}|\sum_a\vec x_a\times\vec q_a
\,\frac{1}{({\cal E}-H_{\rm el})'}\,
\sum_{b}\frac{\vec x_b\times\vec q_b}{x_b^3}\,
\frac{1}{({\cal E}-H_{\rm el})'}
\,\stackrel{\leftrightarrow}{H_{\rm n}}|\phi_{\rm el}\rangle\biggr], \\
\sigma^{(1)}_{\rm d} &=& -\frac{\alpha^2}{3\,g_A}\,\langle\phi_{\rm el}\biggl|
\frac{(g_A-1)}{m_A}\,\sum_{b}\frac{\vec x_b}{x_b^3}\cdot
(\vec x_{\rm el}+\vec R) -\frac{1}{m_B\,R}
\biggr|\phi_{\rm el}\rangle, \\
\sigma^{(1)}_{\rm s} &=& 
\frac{\alpha^2}{3\,g_A}\,\langle\phi_{\rm el}|\sum_a\vec x_a\times\vec q_a\,
\frac{1}{({\cal E}-H_{\rm el})'}
\biggl\{
\frac{1}{m_A}\,\sum_{b}\frac{\vec x_b}{x_b^3}\times
\Bigl[\vec P+(g_A-1)\,\vec q_{\rm el} \Bigr]
\nonumber \\ &&
-\frac{\vec R}{R^3}\times
\biggl[\frac{g_A}{m_B}\,\vec P+
       \frac{(g_A-1)}{m_A}\,\bigl(\vec P-\vec q_{\rm el}\bigr)
       \biggr]\biggr\}|\phi_{\rm el}\rangle,\\
\sigma^{(1)}_{\rm l} &=&\frac{\alpha^2}{3}\,\langle\phi_{\rm el}|
\biggl[ 
\frac{1}{m_A}\,\bigl(\vec x_{\rm el}+\vec R\bigr)\times
\bigl(2\,\vec P-\vec q_{\rm el}\bigr)
-\frac{1}{m_{\rm n}}\,\vec R\times\vec P\biggr]
%\nonumber \\ &&
\frac{1}{({\cal E}-H_{\rm el})'}\,
\sum_b\frac{\vec x_b\times\vec q_b}{x_b^3}|\phi_{\rm el}\rangle. 
\nonumber \\
\end{eqnarray}
These formulae can be further simplified by shifting the reference frame
to the geometrical center and by using gerade symmetry of the ground
electronic state of H$_2$ and isotopomers.

\section{Summary}
We have presented a general approach to finite nuclear mass corrections
in molecular properties which is based on the nonadiabatic perturbation theory \cite{nonad}. 
These corrections were represented in terms of electronic matrix elements
averaged with the nuclear wave function, similarly to that in the adiabatic
approximation.
We obtained formulae for nonadiabatic relativistic corrections
which can be used to perform accurate calculations of dissociation
and rovibrational energies. Currently the accuracy of theoretical predictions
in H$_2$ \cite{jeziorsk} are limited by these not well known effects.
Similarly, we obtained formulae for the nonadiabatic corrections to
the transition electric dipole moment and the electric dipole polarizability.
They can be used for the comparison with precise measurements of 
polarizabilities, for example in such complicated system as excited 
vibrational states of the water molecule \cite{h20a,h20b}.
Finally, we presented nonadiabatic corrections to the magnetic shielding,
which are important for molecules involving hydrogen or deuterium,
where the finite nuclear mass  significantly~$\sim 10^{-3}$
affects the magnetic shielding.

\section*{Acknowledgments}
Author wishes to acknowledge interesting discussion with B. Jeziorski and J. Komasa. 
This work was supported by NIST through Precision Measurement Grant PMG 60NANB7D6153.
%\newpage


\begin{thebibliography}{99}
\bibitem{gyro} The definition of the g-factor assumed here
               is different from the standard one by the use of 
               the actual charge $e$ and the mass $m$ of the corresponding
               particle, namely the coupling of the spin to the magnetic 
               field is $-e\,g/(2\,m)\,\vec s\cdot\vec B$
\end{thebibliography}
\end{document}